\documentclass[proceedings]{JHEP} 

\usepackage{epsfig,axodraw}                   

\newcommand{\ee}{e^+e^-}
\newcommand{\tb}{\tan\beta}

\newcommand{\cl}{$\tilde{\chi}_{1}^\pm$}

\conference{Corfu Summer Institute on Elementary Particle Physics, 1998}

\title{Determining SUSY Parameters from Chargino Pair Production}

\author{Jan Kalinowski\\
        Instytut Fizyki Teoretycznej, Ho\.za 69, 00681 Warszawa, Poland\\
        E-mail: \email{kalino@fuw.edu.pl}}

\abstract{
A procedure to determine the chargino 
mixing angles and, subsequently,  the fundamental SUSY parameters
$M_2, \mu$ and $\tb$ by measurements of the total cross section and 
the spin  correlations in $\ee$ annihilation to $\tilde{\chi}_1^+ 
\tilde{\chi}_1^-$ 
chargino pairs is discussed.}
\dedicated{IFT/99-10}

\begin{document} 
\section{Introduction} Despite the lack of direct experimental
evidence for supersymmetry (SUSY), the concept of symmetry between
bosons and fermions \cite{susy} has so many attractive features that
the supersymmetric extension of the Standard Model is widely
considered as a most natural scenario. SUSY ensures the cancellation
of quadratically divergent corrections from scalar and fermion
loops and thus stabilizes the Higgs boson mass in the desired range of
order $10^2$ GeV, predicts the renormalized electroweak mixing angle
$\sin^2 \theta_W$ in striking agreement with the measured value, and
provides the opportunity to generate the electroweak symmetry breaking
radiatively.

Supersymmetry predicts the quarks and leptons to have scalar partners,
called squarks and sleptons, the gauge bosons to have fermionic
partners, called gauginos. In the Minimal Supersymmetric Standard
Model (MSSM) \cite{mssm} two Higgs doublets with opposite
hypercharges, and with their superpartners: higgsinos, are required to
give masses to the up and down type fermions and to ensure anomaly
cancellation. The higgsinos and electroweak gauginos mix; the mass
eigenstates are called charginos $\tilde{\chi}_{1,2}^\pm$ and
neutralinos $\tilde{\chi}_{1,2,3,4}^0$ for electrically charged and
neutral states, respectively. Actually in many supersymmetric models
the lighter chargino states \cl \ are expected to be the lightest
observable supersymmetric particles and they may play an important
role in the first direct experimental evidence for supersymmetry.

The doubling of the spectrum of states in the MSSM together with our
ignorance on the dynamics of the supersymmetry breaking mechanisms
gives rise to a large number of unknown parameters. Even with the
$R$-parity conserving and CP-invariant SUSY sector, which we will
assume in what follows, in total more than 100 new parameters are
introduced.  This number of parameters can be reduced by additional
physical assumptions.  In the literature several theoretically
motivated scenarios have been considered. The most radical reduction
is achieved by embedding the low--energy supersymmetric theory into a
grand unified (SUSY-GUT) framework called mSUGRA.  The mSUGRA is fully
specified at the GUT scale by a common gaugino mass $m_{1/2}$, a
common scalar mass $m_0$, a common trilinear scalar coupling $A_G$, the ratio
$\tan\beta=v_2/v_1$ of the vev's of the two neutral Higgs fields, and
the sign of the Higgs mass parameter $\mu$. All the couplings, masses
and mixings at the electroweak scale are then determined by the RGEs
\cite{msugra}.  It turns out, however, that the interpretation of
experimental data and derived limits on sparticle masses and their
couplings strongly depends on the adopted scenario.  Therefore it is
important to develop strategies to measure all low-energy SUSY
parameters independently of any theoretical assumptions.  The
experimental program to search for and explore SUSY at present and
future colliders should thus include the following points:
\begin{itemize}
\item[(a)] discover supersymmetric particles and measure their quantum
   numbers to prove that they are {\it  superpartners} of standard
   particles,
\item[(b)]
   determine the low-energy Lagrangian parameters,
\item[(c)] verify the relations among them in order to distinguish
   between various SUSY models.
\end{itemize}
If SUSY is realized in Nature, it will be a matter of days for the
future $e^+e^-$ linear colliders (LC) to discover the kinematically
accessible supersymmetric particles.  Once they are discovered, the
priority will be to measure the low-energy SUSY parameters
independently of theoretical prejudices and then check whether the
correlations among parameters, if any, support a given theoretical
framework \cite{epi99}, like SUSY-GUT relations.  Particularly in this
respect the $\ee$ linear colliders are indispensable tools, as has
been demonstrated in a number of dedicated workshops \cite{LCwork}.
In my talk I will concentrate on the first phase of LC, $i.e.$ when
only a limited number of supersymmetric particles can kinematically be
produced.  In contrast to earlier analyses \cite{R6, R6A}, I will not
elaborate on global chargino/neutralino fits but rather attempt to
explore the event characteristics to isolate the chargino sector.  The
analysis will be based strictly on low--energy supersymmetry.  We will
see that even if the lightest chargino states \cl \ are, before the
collider energy upgrade, the only supersymmetric states that can be
explored experimentally in detail, some of the fundamental SUSY
parameters can be reconstructed from the measurement of char\-gino mass
$m_{\tilde{\chi}_1^\pm}$, the total production cross section and the
chargino polarization in the final state. Beam polarization is helpful
but not necessarily required.

The results presented here have been obtained in collaboration with
S.Y.~Choi, A.~Djouadi, H. Dreiner and P.~Zerwas \cite{choi1}. For an
alternative way of determining the SUSY parameters from the
measurement of (some) chargino and neutralino masses, see \cite{KM}.

\section{Chargino masses and couplings}
The superpartners of the $W$ boson and 
charged Higgs boson, $\tilde{W}^\pm$ and $\tilde{H}^\pm$, 
necessarily  mix since the mass matrix (in the
$\{\tilde{W}^-,\tilde{H}^-\}$ basis)  \cite{mssm} 
\begin{eqnarray}
{\cal M}_C=\left(\begin{array}{cc}
                M_2                &      \sqrt{2}m_W c_\beta  \\
             \sqrt{2}m_W s_\beta   &             \mu   
                  \end{array}\right)
\label{eq:mass matrix}
\end{eqnarray}
is nondiagonal ($c_\beta=\cos\beta$, $s_\beta=\sin\beta$). 
It  is expressed in terms of 
the fundamental supersymmetric parameters: the gaugino
mass $M_2$, the Higgs mass parameter $\mu$, and $\tb$. 
Two different matrices acting on the left-- and right--chiral
$(\tilde{W},\tilde{H})$ states are needed to diagonalize the
asymmetric mass matrix (\ref{eq:mass matrix}).  The  (positive)
eigenvalues are given by 
\begin{eqnarray}
m^2_{\tilde{\chi}^\pm_{1,2}}
  ={\textstyle \frac{1}{2}}\left[M^2_2+\mu^2+2m^2_W
    \mp \Delta \, \right]
\end{eqnarray}
where
\begin{eqnarray}
\Delta=&&[(M^2_2+\mu^2+2m^2_W)^2 \nonumber\\
&&-4(M_2\mu-m^2_W\sin 2\beta)^2]^{1/2}
\end{eqnarray} 
The left-- and right--chiral components of the corresponding 
chargino mass eigenstate 
$\tilde{\chi}^-_1$ are related to the wino and higgsino components
in the following way  
\begin{eqnarray}
&&\tilde{\chi}^-_{1L}=\tilde{W}^-_L\cos\phi_L
                     +\tilde{H}^-_{1L}\sin\phi_L \nonumber\\
&&\tilde{\chi}^-_{1R}=\tilde{W}^-_R\cos\phi_R
                     +\tilde{H}^-_{2R}\sin\phi_R 
\end{eqnarray}
with the rotation angles  given by
\begin{eqnarray}
&&\cos 2\phi_L=-(M_2^2-\mu^2-2m^2_W\cos 2\beta)/\Delta ~~~~~~~~
\nonumber\\ 
&&\cos 2\phi_R=-(M_2^2-\mu^2+2m^2_W\cos 2\beta)/\Delta
\label{mixing}
\end{eqnarray}
As usual, we take $\tan\beta$ positive, $M_2$ positive and $\mu$ of either
sign.

The  angles $\phi_L$ and $\phi_R$ determine the 
$\tilde{\chi}\tilde{\chi}Z$ and the $\tilde{\chi}\tilde{\nu}e$ couplings:
\begin{eqnarray}
&&\langle\tilde{\chi}^-_{1L}|Z|\tilde{\chi}^-_{1L}\rangle 
  = -\frac{e}{4s_W c_W} (4s_W^2 - 3-\cos 2\phi_L) \nonumber\\
&&\langle\tilde{\chi}^-_{1R}|Z|\tilde{\chi}^-_{1R}\rangle 
  = -\frac{e}{4s_W c_W} (4s_W^2-3-\cos 2\phi_R) \nonumber\\
&&\langle\tilde{\chi}^-_{1R}|\tilde{\nu}|e^-_L\rangle 
  =-\frac{e}{s_W}\cos\phi_R
\label{eq:vertex}
\end{eqnarray}
where $s_W^2 =1-c_W^2 \equiv \sin^2\theta_W$. The coupling to the
higgsino component, being proportional to the electron mass, has been
neglected in the sneutrino vertex. The photon--chargino vertex is diagonal,
$i.e.$ it is independent of the mixing angles:
\begin{eqnarray}
\langle\tilde{\chi}^-_{1L,R}|\gamma|\tilde{\chi}^-_{1L,R}\rangle =e 
\end{eqnarray}
The chargino couplings, and therefore mixing angles, are
physical observables and they can be measured. Their knowledge 
together with
the measurement of the chargino masses is sufficient to  determine the
fundamental supersymmetric parameters $M_2$, $\mu$ and $\tan\beta$. 

\section{Chargino production and decay}
Charginos are produced in $e^+e^-$ collisions, either in diagonal or in mixed 
pairs. Here we will consider the 
diagonal pair production 
of the lightest chargino $\tilde{\chi}_1^\pm$ in $e^+e^-$ collisions, 
\begin{eqnarray}
e^+e^- \ \rightarrow \ \tilde{\chi}^+_1 \ \tilde{\chi}^-_1 
\label{eq:eexx}
\end{eqnarray}
assuming the second chargino $\tilde{\chi}_2^\pm$ too heavy to
be produced in the first phase of $e^+e^-$ linear colliders.  If the
chargino production angle could be measured unambiguously on an
event-by-event basis, the chargino couplings could be extracted
directly from the angular dependence of the cross section.  However,
charginos are not stable.  We will assume that they decay into the
lightest neutralino $\tilde{\chi}^0_1$, which is taken to be stable,
and a pair of quarks and antiquarks or charged leptons and neutrinos.
Since two neutral particles $\tilde{\chi}^0_1$ escape undetected, it
is not possible to reconstruct the events unambiguously. In particular
the production angle $\Theta$ cannot be reconstructed for a given
event.  The distribution of observed final state quark jets or leptons
is given by a convolution of the chargino production and decay
processes.  Therefore we will consider the production of polarized
charginos and their subsequent decays.  It has to be stressed that the
spin correlations between production and decay have to be properly
taken into account \cite{R6A,choi1}.

\subsection{Polarized chargino production}
The process $e^+e^-\rightarrow\tilde{\chi}^+_1\tilde{\chi}^-_1$ is generated
by the three mechanisms shown in Fig.~1: $s$--channel $\gamma$ and $Z$
%
\FIGURE[htb]{\epsfig{file=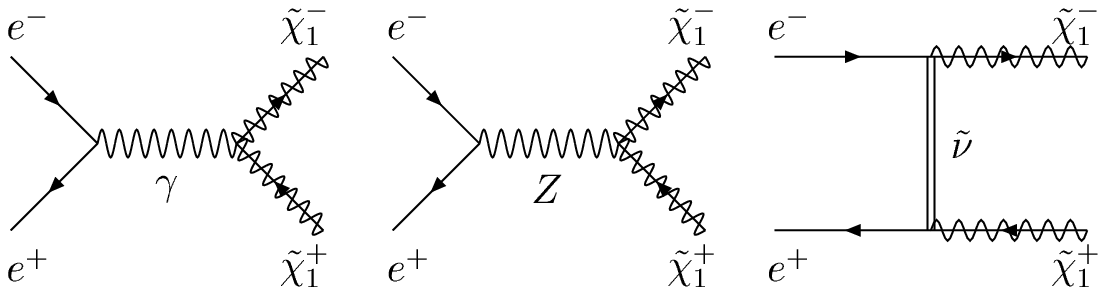,width=7cm}\caption{
{\it The three mechanisms contributing to the production of 
            diagonal chargino  pairs 
            $\tilde{\chi}^+_1 \tilde{\chi}^-_1$ in $\ee$ annihilation.}}}
exchanges, and $t$--channel $\tilde{\nu}$ exchange.
After a Fierz transformation of the $\tilde{\nu}$--exchange
term, the amplitude
\begin{eqnarray}
T= &&\frac{e^2}{s}Q_{\alpha\beta}
   [\bar{v}(e^+)  \gamma_\mu P_\alpha u(e^-)] \nonumber \\
  &&\times [\bar{u}(\tilde{\chi}^-_1) \gamma^\mu P_\beta 
               v(\tilde{\chi}^+_1)]
\label{eq:production amplitude}
\end{eqnarray}
can be expressed in terms of four 
bilinear charges, classified according to the chiralities 
of the associated lepton ($\alpha=L,R$) and chargino ($\beta=L,R$) currents
\begin{eqnarray}
Q_{L\beta}=&&1+ \frac{D_Z}{8s_W^2 c_W^2}(2s_W^2 -1) 
         (4s_W^2 -3-\cos 2\phi_\beta) 
         \nonumber\\ 
      && {} + \delta_{\beta,R}\frac{D_{\tilde{\nu}}}{4s_W^2} 
            (1+\cos 2\phi_R) \nonumber\\
Q_{R\beta}=&&1+\frac{D_Z}{4c_W^2} (4s_W^2 -3-\cos 
          2\phi_\beta) 
\end{eqnarray}
The $\tilde{\nu}$ exchange affects only the $LR$ charge while all
other charges are built up by $\gamma$ and $Z$
exchanges. $D_{\tilde{\nu}}$ denotes the sneutrino propagator
$D_{\tilde{\nu}} = s/(t- m_{\tilde{\nu}}^2)$, and $D_Z$ the $Z$
propagator $D_Z=s/(s-m^2_Z+im_Z\Gamma_Z)$. Above the $Z$ pole the
non--zero width can be neglected so that the charges are real.

The $\tilde{\chi}^-_1$ and $\tilde{\chi}^+_1$ helicities are in
general not correlated due to the non--zero masses of the particles;
amplitudes with equal $\tilde{\chi}^\pm_1$ helicities vanish only
$\propto m_{\tilde{\chi}^\pm_1} /\sqrt{s}$ for asymptotic energies.
Denoting the electron helicity by the first index
$\sigma$ and  the $\tilde{\chi}^-_1$ and $\tilde{\chi}^+_1$ helicities by
the remaining two indices $\lambda$ and $\bar{\lambda}$,  
the helicity amplitudes  $T(\sigma;\lambda \bar{\lambda})=
2\pi\alpha{\cal A}_{\sigma;\lambda
\bar{\lambda}}$ can be expressed as 
functions of bilinear chirality charges  \cite{R8}
\begin{eqnarray}
&& {\cal A}_{+;++}
   =-\sqrt{\beta_-\beta_+}(Q_{RR}+Q_{RL})\sin\Theta \nonumber\\
&& {\cal A}_{+;+-} 
   =-(\beta_+Q_{RR}+\beta_-Q_{RL})(1+\cos\Theta) \nonumber\\
&& {\cal A}_{+;-+} 
   =+(\beta_-Q_{RR}+\beta_+Q_{RL})(1-\cos\Theta) \nonumber\\
&& {\cal A}_{+;--}
   =+\sqrt{\beta_-\beta_+}(Q_{RR}+Q_{RL})\sin\Theta \nonumber \\ 
&& {\cal A}_{-;++}
   =-\sqrt{\beta_-\beta_+}(Q_{LR}+Q_{LL})\sin\Theta \nonumber\\
&& {\cal A}_{-;+-}
   =+(\beta_+Q_{LR}+\beta_-Q_{LL})(1-\cos\Theta) \nonumber\\
&& {\cal A}_{-;-+}
   =-(\beta_-Q_{LR}+\beta_+Q_{LL})(1+\cos\Theta) \nonumber\\
&& {\cal A}_{-;--}
   =+\sqrt{\beta_-\beta_+}(Q_{LR}+Q_{LL})\sin\Theta 
\label{eq:helicity amplitude}
\end{eqnarray}
where $\beta_{\pm}=1\pm\beta$, and  
$\beta=(1-4m^2_{\tilde{\chi}^\pm_1}/s)^{1/2}$ is the $\tilde{\chi}^\pm_1$
velocity in the c.m.~frame. 

The unpolarized differential cross section for 
$e^+e^-\rightarrow\tilde{\chi}^+_1\tilde{\chi}^-_1$ is obtained by 
averaging/sum\-ming over 
the electron/chargino helicities. It can  be written as  
\begin{eqnarray}
\frac{{\rm d}\sigma}{{\rm d}\cos\Theta}
 &&=\frac{\pi\alpha^2}{2 s} \beta 
  \{(1+\beta^2\cos^2\Theta)Q_1\nonumber \\
&&+(1-\beta^2)Q_2
         +2\beta\cos\Theta Q_3\}~~~~~
\label{eq:cross section}
\end{eqnarray}
where the quartic charges $Q_i$  are given in terms of
bilinear charges as follows \cite{R7}
\begin{eqnarray}
&&    Q_1={\textstyle \frac{1}{4}}(|Q_{RR}|^2+|Q_{LL}|^2
                          +|Q_{RL}|^2+|Q_{LR}|^2) \nonumber\\
&&    Q_2={\textstyle \frac{1}{2}}{\rm Re}(Q_{RR}Q^*_{RL}
                                  +Q_{LL}Q^*_{LR}) \\
&&    Q_3={\textstyle \frac{1}{4}}(|Q_{RR}|^2+|Q_{LL}|^2
                          -|Q_{RL}|^2-|Q_{LR}|^2) \nonumber
\end{eqnarray}

%
\FIGURE[htb]{\epsfig{file=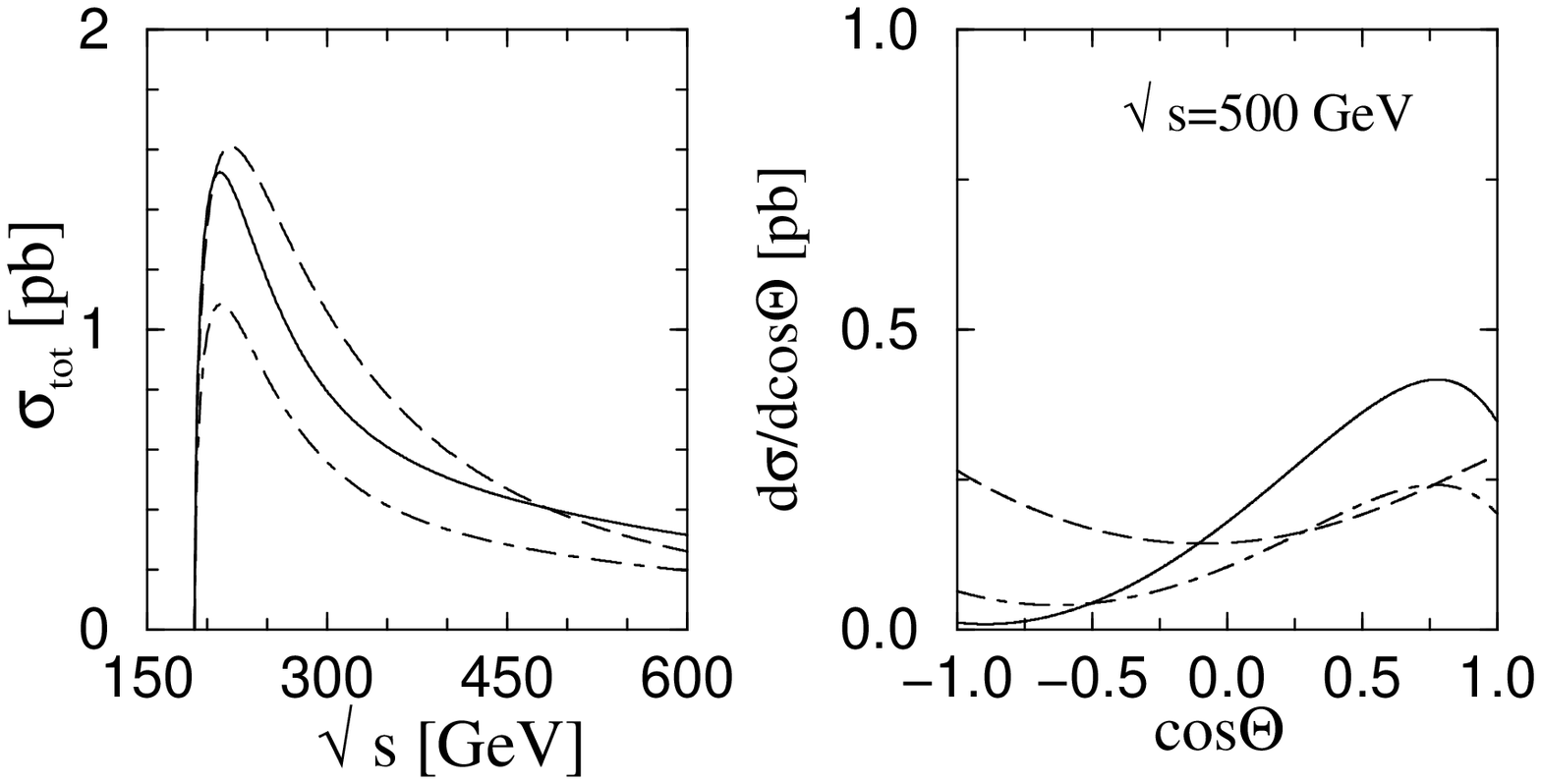,width=7.2cm}\caption{ {\it Left panel:
  the total cross section as a function of the c.m.energy.  Right
  panel: the angular distribution as a function of the production
  angle for $\protect\sqrt{s}=500$ GeV. In both panels:
  $m_{\tilde{\nu}}=200$ GeV, solid lines for the gaugino, dashed lines
  for the higgsino and dot-dashed lines for the mixed cases, see
  eq.~\ref{eq:parameter} [taken from \cite{choi1}].}}}
The total production cross section as a function of the c.m.energy 
and the angular distribution  at $\sqrt{s}=500$ GeV are
shown in Fig.~2 for a representative set of ($M_2,\mu$) parameters;
the sneutrino mass has been taken $m_{\tilde{\nu}}=200$ GeV. 
The parameters are chosen in the higgsino region $M_2 \gg |\mu|$, the gaugino 
region $M_2 \ll |\mu|$ and in the mixed region $M_2 \sim |\mu|$ for 
$\tan\beta=2$ as (in GeV)
\begin{eqnarray}
\begin{array}{ll}
{\rm gaugino}:  & (M_2,\mu)=(81,\, -215)\\
{\rm higgsino}: & (M_2,\mu)=(215,\, -81)\\
{\rm mixed}   : & (M_2,\mu)=(92,\,   -93)
\label{eq:parameter}
\end{array}
\end{eqnarray}
for which the light chargino mass $m_{\tilde{\chi}^\pm_1}$ is
approximately $95$ GeV. The sharp rise of the production cross section
in Fig.~2 allows us to measure the chargino mass
$m_{\tilde{\chi}^\pm_1}$ very precisely.  As is well-known, the
sneutrino exchange leads to a strong destructive interference for the
gaugino and mixed regions, while the dependence of the cross section
on $m_{\tilde{\nu}}$ decreases as the higgsino component of the
chargino increases.  Prior or simultaneous determination of
$m_{\tilde{\nu}}$ is therefore necessary to determine the other SUSY
parameters.

In general charginos are produced with non-zero polarization.  The key
point in our analysis in Sect.~4 will be to exploit the partial
information on the chargino polarizations that can be obtained from
the distribution of the chargino decay products.  Therefore we define the
$\tilde{\chi}^-_1$ polarization vector  $\vec{\cal P}=({\cal
P}_T,{\cal P}_N, {\cal P}_L)$ in the $\tilde{\chi}^-_1$ rest frame.
${\cal P}_L$ denotes the component parallel to the $\tilde{\chi}^-_1$
flight direction in the c.m. frame, ${\cal P}_T$ the transverse
component in the production plane, and ${\cal P}_N$ the component
normal to the production plane. The three components can be written in
terms the helicity amplitudes as 
\begin{eqnarray}
 {\cal P}_L   = { \frac{1}{4{\cal N}}}\sum_{\sigma=\pm} (
          |{\cal A}_{\sigma;++}|^2+|{\cal A}_{\sigma;+-}|^2 \rule{2cm}{0cm} 
                      \nonumber \\
      -|{\cal A}_{\sigma;-+}|^2-|{\cal A}_{\sigma;--}|^2) \rule{2cm}{0cm}
 \\
 {\cal P}_T =\frac{1}{2{\cal N}}\sum_{\sigma=\pm} {\rm Re}
          ( {\cal A}_{\sigma;++}{\cal A}_{\sigma;-+}^* 
          +{\cal A}_{\sigma;--}{\cal A}_{\sigma;+-}^*)
            \nonumber\\
 {\cal P}_N ={ \frac{1}{2{\cal N}}}\sum_{\sigma=\pm}{\rm Im}
          ( {\cal A}_{\sigma;--}{\cal A}_{\sigma;+-}^* 
          -{\cal A}_{\sigma;++}{\cal A}_{\sigma;-+}^*)          
\nonumber 
\label{eq:polarization vector}
\end{eqnarray}
with the normalization
\begin{eqnarray}
{\cal N} ={ \frac{1}{4}}
\sum_{\sigma=\pm}  (|{\cal A}_{\sigma;++}|^2+|{\cal A}_{\sigma;+-}|^2 
    \rule{2cm}{0cm}    \nonumber \\
      +|{\cal A}_{\sigma;-+}|^2+|{\cal A}_{\sigma;--}|^2) 
\rule{1.5cm}{0cm}
\end{eqnarray}
The normal component ${\cal P}_N$ can only be generated by complex
production amplitudes.  Neglecting small higher order loop effects and
the small $Z$--width effect, the normal $\tilde{\chi}^-_1$ and
$\tilde{\chi}^+_1$ polarizations are zero since the $\tilde{\chi}_1
\tilde{\chi}_1 \gamma$ and $\tilde{\chi}_1 \tilde{\chi}_1 Z$ vertices
are real (even for non-zero phases in the chargino mass matrix) and the
$\tilde{\nu}$--exchange amplitude is real too. The CP--violating phases
will change the chargino mass and the mixing angles \cite{choi1,choi2} but
they do not induce complex charges in the production amplitudes of the
diagonal char\-gino pairs.

\subsection{Decay of polarized charginos}

We will assume that charginos decay into the lightest 
neutralino $\tilde{\chi}^0_1$ and a pair of quarks and antiquarks or
leptons and neutrinos: $\tilde{\chi}^\pm_1\rightarrow
\tilde{\chi}^0_1+f\bar{f}'$.  The decay can proceed via the exchange
of the $W$, squarks or sleptons; the exchange of the charged Higgs
bosons can be neglected for light fermions in the final state. The
corresponding diagrams are shown in Fig.~3 for the decay into quark
pairs.
%
\FIGURE[ht]{\epsfig{file=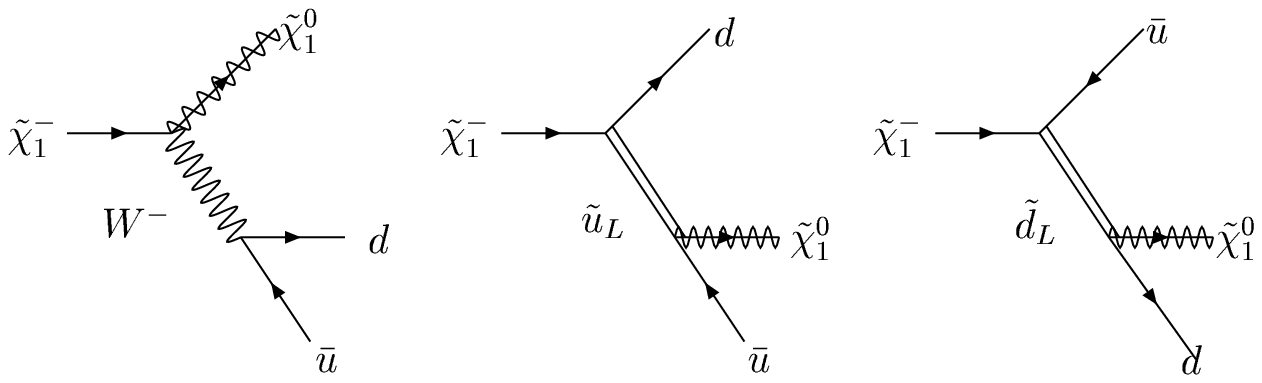,width=7cm}\caption{{\it Chargino
               decay mechanisms; the exchange of the charged Higgs
               boson is  neglected.}}}
After a  Fierz transformation of the $\tilde{q}$--exchange
contributions, the decay amplitude
$\tilde{\chi}^-_1\rightarrow\tilde{\chi}^0_1d\bar{u}$ may be written
as:
\begin{eqnarray}
{\cal D}
 =&&\frac{e^2}{2 \sqrt{2}s_W^2}
  [\bar{u}(d)\gamma^\mu P_L v(\bar{u})]\nonumber \\
&&\times [\bar{u}(\tilde{\chi}^0_1)\gamma_\mu
                   (F_LP_L+F_RP_R) u(\tilde{\chi}^-_1)]~~~
\end{eqnarray}
with 
\begin{eqnarray}
F_L&&=D_W(2 N_{12} \cos\phi_L + \sqrt{2} N_{13}\sin\phi_L)\nonumber \\
      &&  {}     + D_{\tilde{d}_L} \cos\phi_L( N_{12}
- 2 Y_q \tan\theta_W N_{11}) \nonumber\\
F_R&&=D_W( 2 N^*_{12} \cos\phi_R- \sqrt{2} N^*_{14}\sin\phi_R) \nonumber \\
      && {}+D_{\tilde{u}_L}\cos\phi_R( N^*_{12}+ 2 Y_q \tan\theta_W N^*_{11})
~~~~~
\end{eqnarray}
where $Y_q=1/6$ is the quark hypercharge and 
$D_W= s'-m^2_W+im_W\Gamma_W$, 
$D_{\tilde{d}_L}=t'-m^2_{\tilde{d}_L}$, 
$D_{\tilde{u}_L}=u'-m^2_{\tilde{u}_L}$. 
Analogous expressions 
apply to decays into lepton pairs with $Y_l=-1/2$.  
The Mandelstam variables $s'$, $t'$, $u'$ 
are defined in terms of the 4--momenta $q_0,q$ and $\bar q$ 
of $\tilde{\chi}_1^0, d$ and $\bar{u}$, 
respectively, as 
$s'=(q+\bar{q})^2$, $t'=(q_0+q)^2$, $u'=(q_0 +\bar{q})^2 $, 
while $N_{ij}$ is the $4\times 4$ matrix rotating  
the current neutralino eigenstates 
$\tilde{B}, \tilde{W}^3,  \tilde{H}^0_1,  \tilde{H}^0_2$ to the mass
eigenstates $\tilde{\chi}^0_1,.., \tilde{\chi}^0_4$.
The neutralino mass matrix ${\cal M}_N$ is given by 
\begin{eqnarray}
 \left[ \begin{array}{cccc}
M_1 & 0 & -m_Z s_W c_\beta & m_Z  s_W s_\beta \\
0   & M_2 & m_Z c_W c_\beta & -m_Z  c_W s_\beta \\
-m_Z s_W c_\beta & m_Z  c_W c_\beta & 0 & -\mu \\
m_Z s_W s_\beta & -m_Z  c_W s_\beta & -\mu & 0
\end{array} \right] \nonumber
\end{eqnarray}
Besides the parameters $M_2, \mu$ and $\tan \beta$, which already
appear in the chargino mass matrix, the only additional parameter in
the neutralino mass matrix is $M_1$\footnote{In GUT models with the
unification of gaugino masses at a high scale, the $M_1$ and $M_2$ are
related by $M_1= \frac{5}{3} \tan^2 \theta_W M_2$.}. In general
supersymmetric models, however, the neutralino sector can potentially
be more complex than in the MSSM with more additional parameters.

In the chargino rest frame, the angular distribution of the neutralino
and the $f_1\bar f_2$ system from the polarized chargino decay
$\tilde{\chi}^\pm_1\rightarrow \tilde{\chi}^0_1+f_1\bar{f}_2$ (summed
over the final state polarizations) is determined by the energy and
the polar angle of the neutralino (or equivalently of the $f_1 \bar
f_2$ system) with respect to the chargino polarization vector.  Let us
define $\theta^*$ ($\bar{\theta}^*$) as the polar angle of the
$f_1\bar{f}_2$ ($\bar{f}_3f_4$) system in the $\tilde{\chi}^-_1$
($\tilde{\chi}^+_1$) rest frame with respect to the original flight
direction in the laboratory frame, and $\phi^*$($\bar{\phi}^*$) the
corresponding azimuthal angle with respect to the production plane.
Taking the $\tilde{\chi}^-_1$ ($\tilde{\chi}^+_1$) 
flight direction as quantization axis
and for the kinematical configuration defined above, the spin--density
matrix $\rho_{\lambda\lambda'}\sim {\cal D}_\lambda {\cal
D}^*_{\lambda'}$ can conveniently be written in the following form
\begin{eqnarray}
 \rho_{\lambda\lambda^\prime}
   =\frac{1}{2}\left(\begin{array}{cc}
         1+\kappa\cos\theta^*         &   \kappa\sin\theta^*{\rm e}^{i\phi^*} 
 \\ \kappa\sin\theta^*{\rm e}^{-i\phi^*} &        1-\kappa\cos\theta^*
                     \end{array}\right)~~~
\end{eqnarray}
for the $\tilde{\chi}^-_1$ decay, 
and similarly for the $\tilde{\chi}^+_1$ with barred quantities. 

%
\FIGURE[ht]{\epsfig{file=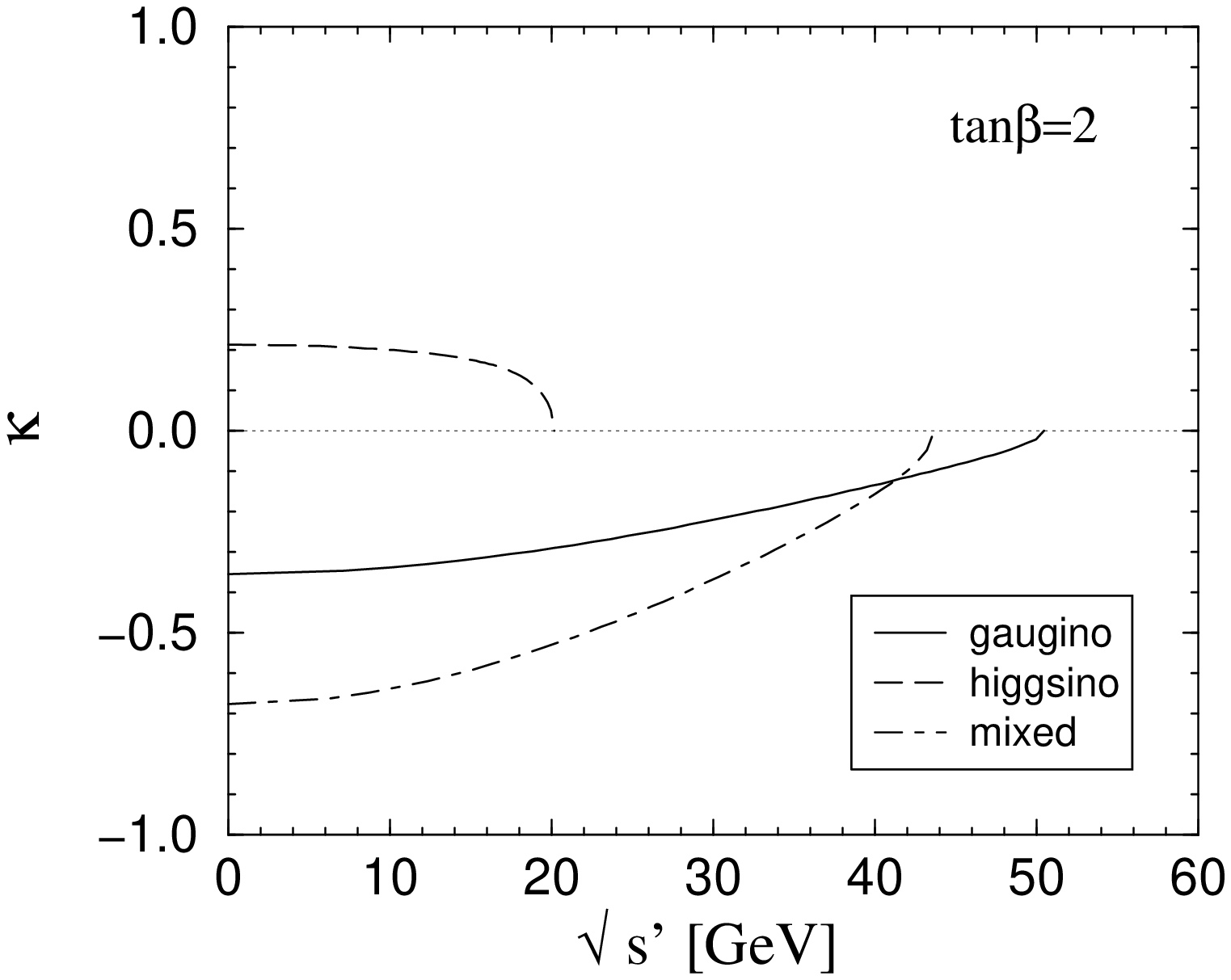,width=7cm}\caption{{\it 
              The polarization analysis--power $\kappa$ as a function of the
             hadron invariant mass $\protect\sqrt{s'}$ for the same 
             set of parameters as 
             for the cross section (\ref{eq:parameter}).}}}

The spin analysis--powers $\kappa$ and $\bar{\kappa}$, depend on the
final $ud$ or $l\nu$ pairs considered in the chargino decays.  Since
left-- and right--chiral form factors $F_L,F_R$ contribute at the same
time, their values are determined by the masses and couplings of all
the particles involved in the decay.  Characteristic examples for
$\kappa$ as a function of the invariant mass of the fermion system are
shown in Fig.~4; the squark masses are set to 300~GeV, and the gaugino
masses are assumed universal at the unification scale.  For our
purposes, however, it will suffice that they are finite.  Let us also
note that neglecting effects from non--zero widths, loops and
CP--noninvariant phases, $\kappa$ and $\bar{\kappa}$ are real, and
$\kappa=-\bar{\kappa}$ for charginos $\tilde{\chi}^\pm_1$ decaying to
charge-conjugated final states.

\subsection{Physical observables}

Charginos decay fast and only correlated production and decay can be
observed experimentally. The matrix element for the physical process
\begin{equation}
e^+e^-\rightarrow\tilde{\chi}^-_1\tilde{\chi}^+_1 \rightarrow
(\tilde{\chi}^0_1f_1\bar{f}_2)(\tilde{\chi}^0_1 \bar{f}_3f_4)~~~~~
\end{equation} 
summed/averaged over the final/initial  state polarizations, is given by 
${\cal M}=\sum T_{\sigma;\lambda\bar{\lambda}}{\cal D}_\lambda
\overline{\cal D}_{\bar{\lambda}}$.  Integrating over the production
angle $\Theta$ and also over the invariant masses of the fermionic
systems $(f_1 \bar{f}_2)$ and $(\bar{f}_3 f_4)$, one can write the
4-fold differential cross section in the following form:
\begin{eqnarray}
\frac{d^4 \sigma }{d\cos \theta^* d\phi^* d\cos \bar{\theta}^*
d\bar{\phi}^*} 
= \frac{\alpha^2 \beta}{124 \pi s} {\cal B}\, \bar{\cal B}
\, \Sigma ~~~~
\end{eqnarray}
where 
${\cal B}=Br(\tilde{\chi}_1^- \to \tilde{\chi}_1^0 f_1 \bar{f}_2)$ and  
${\bar{\cal B}}=Br(\tilde{\chi}_1^+ \to \tilde{\chi}_1^0 \bar{f}_3 f_4)$. 
The scaled differential
cross section 
$\Sigma= \Sigma(\theta^*, \phi^*, \bar{\theta}^*, \bar{\phi}^*)$
is a product of the production helicity amplitudes
${\cal
A}_{\sigma;\lambda\bar{\lambda}}$ and the spin density matrices for the
chargino decays $\rho_{\lambda\lambda'}$\begin{eqnarray}
\Sigma=\sum_{\lambda\bar{\lambda}}\sum_{\lambda'\bar{\lambda}'}\sum_\sigma
       {\cal A}_{\sigma;\lambda\bar{\lambda}}
       {\cal A}^*_{\sigma;\lambda'\bar{\lambda}'}\, 
       \rho_{\lambda\lambda'}\bar{\rho}_{\bar{\lambda}\bar{\lambda}'}
\end{eqnarray}

The quantity $\Sigma$ can be decomposed
into sixteen independent angular parts
\begin{eqnarray}
\Sigma&=&\Sigma_{\rm un}+\cos\theta^*\kappa{\cal P}
       +\cos\bar{\theta}^* \bar{\kappa}\bar{\cal P}\nonumber\\
     &&+\cos\theta^*\cos\bar{\theta}^*\kappa\bar{\kappa}{\cal Q}\nonumber\\
     &&+\sin\theta^*\sin\bar{\theta}^*\cos(\phi^*+\bar{\phi}^*)
        \kappa\bar{\kappa}{\cal Y}\nonumber\\
     &&+\sin\theta^*\sin\bar{\theta}^*\sin(\phi^*+\bar{\phi}^*)
        \kappa\bar{\kappa}\bar{\cal Y}\nonumber\\
     &&+\sin\theta^*\cos\phi^*\kappa{\cal U}
       +\sin\theta^*\sin\phi^*\kappa\bar{\cal U}\nonumber\\
     &&+\sin\bar{\theta}^*\cos\bar{\phi}^*\bar{\kappa}{\cal V}
       +\sin\bar{\theta}^*\sin\bar{\phi}^*\bar{\kappa}\bar{\cal V}\nonumber\\
     &&+\sin\theta^*\cos\bar{\theta}^*
                 (\cos\phi^*\kappa\bar{\kappa}{\cal W}
                 +\sin\phi^*\kappa\bar{\kappa}\bar{\cal W}) \nonumber\\
     &&+\cos\theta^*\sin\bar{\theta}^*
                 (\cos\bar{\phi}^*\kappa\bar{\kappa}{\cal X}
                 +\sin\bar{\phi}^*\kappa\bar{\kappa}\bar{\cal X})\nonumber\\
     &&+\sin\theta^*\sin\bar{\theta}^*
           (\cos(\phi^*-\bar{\phi}^*)\kappa\bar{\kappa}{\cal Z}\nonumber \\
     &&+\sin\theta^*\sin\bar{\theta}^*
           (\sin(\phi^*-\bar{\phi}^*)\kappa\bar{\kappa}\bar{\cal Z} 
\label{eq:sigma}
\end{eqnarray}
where the sixteen coefficients are combinations of helicity amplitudes,
corresponding to the unpolarized cross section ($\Sigma_{\rm un}$), 
$2 \times 3$
polarization components (${\cal P,U,V},\bar{\cal P},\bar{\cal U}, 
\bar{\cal V}$)
and $3 \times 3$ spin--spin correlations (the remaining ones) 
multiplied by the spin-analysis power factors $\kappa$ and
$\bar{\kappa}$.  
In  the CP-invariant theory, and neglecting loops and the width of 
the $Z$--boson for high 
energies, the six functions 
$\bar{\cal U}, \bar{\cal V}, \bar{\cal W}, \bar{\cal X},
\bar{\cal Y}, \bar{\cal Z}$ can be discarded. Moreover, from 
CP--invariance
\begin{eqnarray}
{\cal A}_
{\sigma; \lambda\lambda'}= (-1)^{\lambda-\lambda'-1} {\cal A}_{\sigma; 
-\lambda'-\lambda}
\end{eqnarray}  
it follows that $\bar{\cal P}=-{\cal P}$, ${\cal U}=-{\cal V}$ and ${\cal 
W}={\cal X}$. The overall topology is therefore determined by seven 
independent functions: 
$\Sigma_{\rm un}$, ${\cal P}$, ${\cal Q}$, ${\cal U}$,
${\cal W}$, ${\cal Y}$ and ${\cal Z}$.

The key point
in our analysis is to exploit the partial information on the chargino
polarizations with which they are produced. 
The $\tilde{\chi}$ polarization vectors
and $\tilde{\chi}$--$\tilde{\chi}$ spin--spin correlation tensors 
can
be determined from the decay distributions of the charginos {\it
independently} of the chargino decay dynamics.

The decay angles $(\theta^*,\phi^*)$ and $(\bar{\theta}^*,\bar{\phi}^*)$,
which are used to measure the $\chi_1^\pm$ chiralities, 
can not be reconstructed completely since there are two invisible
neutralinos in the final state.
However, the longitudinal components and the inner product of the transverse
components can be reconstructed from the momenta measured in the laboratory
frame 
\begin{eqnarray}
&&\cos\theta^*= ({E}-{\gamma}E^*)/{\beta {\gamma}|\vec{p}^*|}\nonumber \\
&&\cos\bar{\theta}^*=
({\bar{E}}-{\gamma}\bar{E}^*)/{\beta {\gamma}|\vec{\bar{p}}^*|} \nonumber\\
&&\sin\theta^*\sin\bar{\theta}^*\cos(\phi^*+\bar{\phi}^*)
 ={\vec{p} \cdot \vec{\bar{p}}}
        /{|\vec{p}^*||\vec{\bar{p}}^*|}~~~~\nonumber \\
&&
+(\gamma E-E^*)
           (\gamma \bar{E}-\bar{E}^*)/{\beta^2\gamma^2  
           |\vec{p}^*||\vec{\bar{p}}^*|}
\end{eqnarray}
where $\gamma=\sqrt{s}/{2m_{\tilde{\chi}^\pm_1}}$. $E (\bar{E})$ and
$E^*(\bar{E}^*)$ are the energies of the two hadronic systems in the
laboratory frame and in the rest frame of the charginos, respectively;
$\vec{p} (\vec{\bar{p}})$ and $\vec{p}^* (\vec{\bar{p}^*})$ are the
corresponding momenta; the angle between the vectors in the transverse
plane is given by $\Delta \phi^* = 2\pi-(\phi^*+\bar{\phi}^*)$ for the
reference frames defined earlier.\footnote{Actually to determine the
kinematical variables, $\cos \theta^*$ etc., the knowledge of
$m_{\tilde{\chi}^0_1}$ is also needed, which can be extracted from the
energy distributions of final state particles, see later.  However, it
must be stressed that the above procedure does not depend on the
details of decay dynamics nor on the structure of (potentially more
complex) neutralino and sfermion sectors.}  Therefore, by means of
kinematical projections the terms in the first three lines of
eq.(\ref{eq:sigma}) can be extracted and three $\kappa$-independent
physical observables, $\Sigma_{un}$, ${\cal P}^2/{\cal Q}$ and ${\cal
P}^2/{\cal Y}$, construc\-ted.  They are unambiguously related to the
properties of the chargino sector, not affected by the neutralinos,
since they are given in terms of the helicity production amplitudes as
follows
\begin{eqnarray}
\Sigma_{\rm un}&=&{\textstyle \int} {\rm d}\cos\Theta \, {\cal N}\nonumber \\
{\cal P}&=&{\textstyle \frac{1}{4}\int}
           {\rm d}\cos\Theta \, {\cal N} {\cal P}_L
               \nonumber\\
{\cal Q}&=&{\textstyle\frac{1}{4}\int} {\rm d}\cos\Theta\sum_{\sigma=\pm}
      (|{\cal A}_{\sigma;++}|^2-|{\cal A}_{\sigma;+-}|^2 \nonumber \\
          &&{}~~~~ -|{\cal A}_{\sigma;-+}|^2+|{\cal A}_{\sigma;--}|^2)
      \nonumber\\
{\cal Y}&=&{\textstyle\frac{1}{2}\int} {\rm d}\cos\Theta\sum_{\sigma=\pm} 
      {\rm Re}({\cal A}_{\sigma;--}{\cal A}_{\sigma;++}^*)~~~
\end{eqnarray}

It is thus possible to study 
the chargino sector in isolation by measuring the mass of the lightest
chargino, the total production cross section and 
the spin(--spin) correlations.

\section{Extraction of SUSY parameters}
The pair production of the lightest chargino $\chi_1^\pm$ is 
characterized by the chargino mass $m_{\tilde{\chi}^\pm_1}$ 
and the two mixing 
angles $\cos 2\phi_{L,R}$. For simplicity we assume that the 
sneutrino mass $m_{\tilde{\nu}}$ is obtained from elsewhere, although
combining the energy variation of the cross section
with the measurement of the spin correlations, the sneutrino mass 
$m_{\tilde{\nu}}$ can be also extracted. 
The three quantities $m_{\tilde{\chi}^\pm_1}$ and $\cos 2\phi_{L,R}$
can be determined from the production cross section and the spin
correlations as follows.  

The mass $m_{\tilde{\chi}^\pm_1}$ can be measured very precisely near
the threshold where the production cross section
$\sigma(e^+e^-\rightarrow \tilde{\chi}^+_1\tilde{\chi}^-_1)$ rises
sharply with the chargino velocity
$\beta$. Alternatively the masses of
chargino and neutralino can be determined by fitting the energy
spectra of the jet-jet final state systems \cite{uli}.  For the
analysis below we assume that the mass of the light chargino has been
measured and $m_{\tilde{\chi}^\pm_1}=95$ GeV

Since the polarization ${\cal P}$ is odd under parity and charge--conjugation, 
it is necessary to identify the chargino electric charges in this case. 
This can be accomplished by making use of the mixed 
leptonic and hadronic decays of the chargi\-no pairs. On the other hand, the
observables ${\cal Q}$ and 
${\cal Y}$ are defined without charge identification so that the 
dominant hadronic decay modes of the charginos can be exploited. 
Suppose that the quantities $\sigma_t$, ${\cal P}^2/{\cal Q}$ and 
${\cal P}^2/{\cal Y}$ have been measured and  are taken to be 
\begin{eqnarray}
&&\sigma_t(e^+e^-\rightarrow\tilde{\chi}^+_1\tilde{\chi}^-_1)=0.37\ \ 
{\rm pb} \nonumber \\
&&\frac{{\cal P}^2}{\cal Q}=-0.24,\; 
\frac{{\cal P}^2}{\cal Y}= -3.66
\label{eq:measured}
\end{eqnarray}
at $\sqrt{s}=500$ GeV. 
These measurements  
can be interpreted as 
contour lines in the plane ($\cos 2\phi_L$, $\cos 2\phi_R$) 
which intersect with large angles so that a high precision in the 
resolution can be achieved. A representative example for the 
determination of $\cos 2\phi_L$ and $\cos 2\phi_R$ based on the values 
in eq.~(\ref{eq:measured}) 
is shown in 
Fig.~5. 
%
\FIGURE[ht]{\epsfig{file=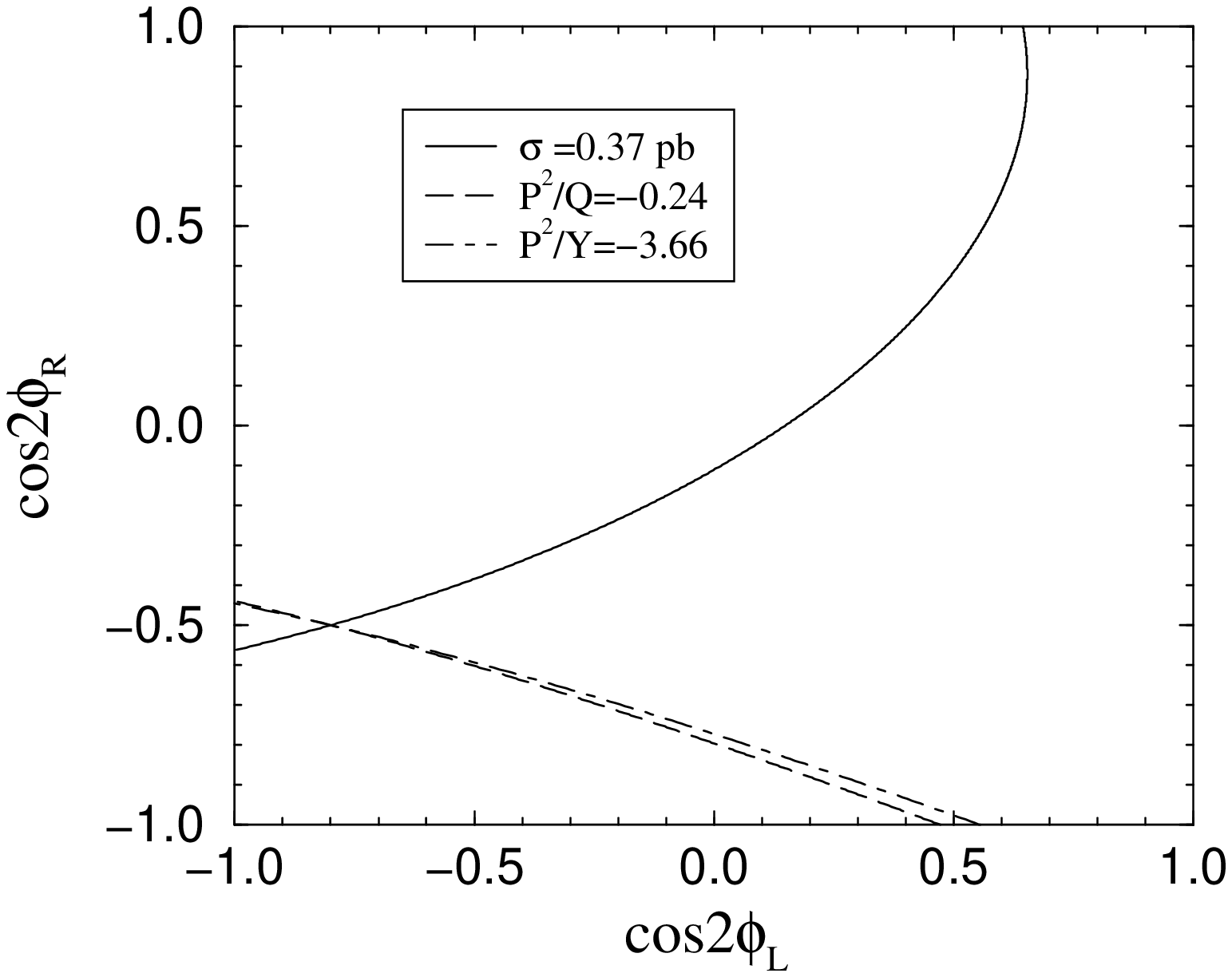,width=7cm}\caption{{\it 
             Contours for the ``measured values''
             (\ref{eq:measured}) of the total 
             cross section (solid line), ${\cal P}^2/{\cal Q}$ (dashed line), 
             and ${\cal P}^2/{\cal Y}$ (dot-dashed line) 
             for $m_{{\chi}^\pm_1}=95$ 
             GeV [$m_{\tilde{\nu}}=250$ GeV].}}}
The three contour lines meet at a single point 
$(\cos 2\phi_L,\cos 2\phi_R)=(-0.58,-0.48)$ for $m_{\tilde{\nu}}=250$ 
GeV; note that
the sneutrino mass can be determined together with the mixing angles 
from the ``measured values" in eq.~(\ref{eq:measured}).

Finally the 
Lagrangian parameters $M_2, \mu$ and $\tan\beta$ can be obtained from 
$m_{\tilde{\chi}^\pm_1}$, 
$\cos 2\phi_L$ and $\cos 2\phi_R$ up to a two-fold ambiguity. It is most
transparently achieved by introducing the two auxiliary quantities
\begin{eqnarray}
p=\cot(\phi_R-\phi_L), \;\;   q=\cot(\phi_R+\phi_L)
\end{eqnarray}
They are expressed in terms of the measured values $\cos 2\phi_L$ and
$\cos 2\phi_R$  up to a discrete ambiguity due to undetermined signs 
 of $\sin 2\phi_L$ and $\sin 2\phi_R$
\begin{eqnarray}
p^2+q^2 &=&\frac{2(\sin^2 2\phi_L+\sin^2 2\phi_R)}{(\cos 2\phi_L-\cos 2
\phi_R)^2}
           \nonumber\\
  pq    &=&\frac{\cos 2\phi_L+\cos 2\phi_R}{\cos 2\phi_L-\cos 2\phi_R}
\nonumber\\
p^2-q^2 &=&\frac{4\sin 2\phi_L\sin 2\phi_R}{(\cos 2\phi_L-\cos 2\phi_R)^2} 
\end{eqnarray}
Solving then eqs.~(\ref{mixing}) for $\tan\beta$ one finds at most 
two possible solutions, and using 
\begin{eqnarray}
M_2&=&{m_W}[(p+q)s_\beta-(p-q)c_\beta]/\sqrt{2}
       \nonumber  \\
\mu&=&{m_W}[(p-q)s_\beta-(p+q)c_\beta]/\sqrt{2}
\label{eq:M2mu}
\end{eqnarray}
we arrive at $\tan\beta$, $M_2$ and $\mu$ up to  a two-fold
discrete ambiguity. For example, taking the ``measured values'' from
eq.~(\ref{eq:measured}), the following results are found 
\cite{choi1}
\begin{eqnarray}
 [\tan\beta; M_2,\mu] =
        \left\{\begin{array}{l}
              [1.06; \;  83{\rm GeV}, \; -59{\rm GeV}] \\
                  { }\\ {}
              [3.33; \;  248{\rm GeV}, \; 123{\rm GeV}]
              \end{array}\right.
\end{eqnarray}
Other sets of ``measured values'' can lead to  a unique solution  
if the other ``possible 
solution'' has a negative  $\tan\beta$. 

To summarize, from the light chargino pair production, the 
measurements of the total production cross section and either of the
angular correlations among the chargino decay products (${\cal
P}^2/{\cal Q}$, ${\cal P}^2/{\cal Y}$),  the physical parameters
$m_{\tilde{\chi}^\pm_1}$, $\cos 2\phi_L$ and $\cos 2\phi_R$ 
are determined
unambiguously. Then the fundamental parameters $\tan\beta$,
$M_2$ and $\mu$ are extracted  up to a two-fold discrete ambiguity. 

If the collider energy is sufficient to produce the two chargino
states in pairs, the above ambiguity can be removed \cite{choi2} by
the measurement of the heavier chargino mass.  With polarized beams
available at the LC, the measurement of the left-right asymmetry
$A_{LR}$ can provide \cite{choi2} an alternative way to extract the
mixing angles (or serve as a consistency check).

\section{Conclusions}

We have discussed how the parameters of the chargino system, the mass
of the light chargino $m_{\tilde{\chi}^\pm_1}$ and the two angles
$\phi_L$ and $\phi_R$, can be extracted from pair production of the
light chargino state in $\ee$ annihilation.  In addition to the
total production cross section, the measurements of angular
correlations among the chargino decay products give rise to two
independent observables which can be measured directly despite of the
two invisible neutralinos in the final state.

{}From the chargino mass $m_{\tilde{\chi}^\pm_1}$ and the two mixing 
parameters 
$\cos 2\phi_{L,R}$, the fundamental supersymmetric parameters $\tan\beta$, 
$M_2$ and $\mu$ can be extracted up to at most a two-fold discrete ambiguity. 
Moreover, from the energy distribution of the final particles
in the decay of the chargino, the mass of the lightest neutralino can be 
measured; this allows us to determine the parameter $M_1$ so that 
also the neutralino mass matrix can be reconstructed
in a model-independent way. 

Although we only
considered real-valued parameters, some of the material presented here
goes through unaltered if phases are allowed \cite{choi1,choi2} even
though extra information will still be needed to determine those
phases.

It should be stressed that the strategy presented here is just at the
theoretical level. More realistic simulations of the experimental
measurements of physical observables and related errors, including
radiative corrections, are needed to assess fully its usefulness.
Nevertheless, if the LC and detectors are built and work as expected,
no doubt that the actual measurements will be better than
anything presented here -- provided supersymmetry is
discovered!

\acknowledgments
I thank the organisers G. Zoupanos, N. Tracas and G. Koutsoumbas for their
warm hospitality at Corfu. 
I would also like to thank my collaborators S.Y. Choi, A. Djouadi, H. Dreiner 
 and P. Zerwas for many valuable discussions. This work has been partially
supported by the KBN grant 2 P03B 052 16.


\begin{thebibliography}{99}

\bibitem{susy} Yu.A. Gol'fand and E.P. Likhtman, \jetpl{13}{1971}{452};
  \\ J.~Wess and B.~Zumino, \npb{70}{1974}{39}; \\
  R.~Haag, J.T.~\L opusza\'nski and M.F.~Sohnius, \npb{88}{1975}{257}.

\bibitem{mssm} For reviews of supersymmetry and the Minimal
   Supersymmetric Standard Model, see H.~Nilles, \prep{110}
   {1984}{1}; \\ H.E.~Haber and G.L.~Kane, \prep{117}{1985}{75}.

\bibitem{msugra} V. Barger, M.S.~Berger and P.~Ohmann, \prd{49}{1994}{4908}.

\bibitem{epi99} J. Kalinowski,  Supersymmetry searches at $e^+e^-$ linear
   colliders, \hepph{9904260}.

\bibitem{LCwork} Proceedings { \em Physics and Experiments with Linear
   Colliders}:\\
  R.~Orava, P.~Eerola, M.~Nordberg (Eds.), Sariselk\"a,
   Finland 1991, World Scientific (1992); \\
  F.A. Harris, S.~Olsen,
   S.~Pakvasa, X.~Tata (Eds.), Waikoloa, Hawaii 1993, World Scientific
   (1994); \\
  A.~Miyamoto, Y.~Fujii, T.~Matsui, S.Iwata (Eds.), Morioka,
   Japan 1995, World Scientific (1996);\\
  E.~Accomando {\it et al.}, LC CDR Report DESY 97-100 
   [\hepph{9705442}], and \prep{299}{1998}{1}.  


\bibitem{R6} A.~Leike, \ijmpa{3}{1988}{2895};\\ 
   M.A.~Diaz and S.F.~King, \plb{349}{1995}{105} and {\bf B373} (1996) 100;\\
   J.L.~Feng and M.J.~Strassler, \prd{51}{1995}{4461} 
   and {\bf D55} (1997) 1326; \\
   G. Moortgat-Pick and
   H.~Fraas, \prd{59}{1999}{015016} [\hepph{9708481}]. 

\def\epjc#1,#2,#3{{\it Eur.\ Phys.\ J.\/ }{\bf C#1} (#2) #3}

\bibitem{R6A} G.~Moortgat-Pick, H.~Fraas, A.~Bartl  and W.~Majerotto,
   \epjc7,1999,113 [\hepph{9804306}];\\
   V. Lafage {it et al.}, Spin and spin-spin correlations 
in chargino pair production at future linear $e^+e^-$ colliders, 
\hepph{9810504}.


\bibitem{choi1} S.Y. Choi, A. Djouadi, H. Dreiner, J. Kalinowski and
P.M. Zerwas, \epjc7,1999,123 [\hepph{9806279}]

\bibitem{KM} J.L. Kneur and G. Moultaka, 
\prd{59}{1999}{015005} [\hepph{9807336}]. 


\bibitem{R8} K.~Hagiwara and D.~Zeppenfeld, \npb{274}{1986}{1}.


\bibitem{R7} L.M.~Sehgal and P.M.~Zerwas, \npb{183}{1981}{417}.

\bibitem{uli} H.U. Martyn, in DESY-ECFA Conceptual LC Design Report,
DESY 1997-048, ECFA 1997-182. For an updated analysis at high
luminosity, see H.U.~Martyn, talk given at the 2nd ECFA-DESY Linear
Collider Workshop, Oxford 1999.

\bibitem{choi2} S.Y. Choi, A. Djouadi,  H.S. Song and P.M. Zerwas,
 Determining SUSY parameters in chargino pair-production in $e^+e^-$ 
collisions, \hepph{9812236}.



\end{thebibliography}
\end{document}